\documentclass[%
reprint,
jmp,
superscriptaddress,
nofootinbib,
amsmath,amssymb,
table
]{revtex4-2}
\usepackage[english]{babel}
\usepackage{graphicx}
\usepackage{dcolumn}
\usepackage{bm}
\usepackage[version=4]{mhchem}
\usepackage{svg}
\usepackage{multirow}
\usepackage{comment}
\usepackage{hyperref}
\usepackage{tabularx}
\usepackage{array}
\usepackage{makecell}
\usepackage{amssymb}
\usepackage{float}
\usepackage{afterpage}
\usepackage[utf8]{inputenc}
\usepackage{braket}
\usepackage{amsmath}
\usepackage{amsfonts}
\usepackage{booktabs} 
\usepackage{titlesec}
\usepackage{dsfont}
\usepackage{pifont}
\usepackage{adjustbox}
\usepackage[normalem]{ulem}
\usepackage{tabularray}
\usepackage[left=0.8in,right=0.8in,top=0.8in,bottom=0.8in]{geometry}
\usepackage{upgreek}
\usepackage{textgreek}
\usepackage[super]{nth}
\usepackage{csquotes}
\usepackage{xcolor}
\usepackage{hhline}

\newcolumntype{Y}{>{\centering\arraybackslash}X}

\renewcommand{\thefootnote}{\fnsymbol{footnote}}

\definecolor{myblue}{RGB}{71,120,207}
\definecolor{mygreen}{RGB}{106,204,100}
\definecolor{myred}{RGB}{213,95,95}

\date{\today}

\makeatletter
\def\@email#1#2{%
 \endgroup
 \patchcmd{\titleblock@produce}
  {\frontmatter@RRAPformat}
  {\frontmatter@RRAPformat{\produce@RRAP{*#1\href{mailto:#2}{#2}}}\frontmatter@RRAPformat}
  {}{}
}%
\makeatother
\graphicspath{{figures/}}
\begin{document}

\preprint{AIP/123-QED}

\makeatletter
\renewcommand\paragraph{%
  \@startsection{paragraph}{4}{\z@}%
    {3.25ex \@plus1ex \@minus.2ex}%
    {-1em}%
    {\normalfont\normalsize\bfseries}%
}
\makeatother

\title{Neural Thermodynamic Integration: Free Energies \\ from Energy-based Diffusion Models}%

\author{Bálint Máté\textsuperscript{*}}
\affiliation{Institute for Theoretical Physics, Heidelberg University, Heidelberg, Germany}
\affiliation{Department of Computer Science, University of Geneva, Carouge, Switzerland}
\affiliation{Department of Physics, University of Geneva, Geneva, Switzerland}
\author{François Fleuret\textsuperscript{\dag}}%
\affiliation{Department of Computer Science, University of Geneva, Carouge, Switzerland}
\author{Tristan Bereau\textsuperscript{\ddag}}%
\affiliation{Institute for Theoretical Physics, Heidelberg University, Heidelberg, Germany}


\begin{abstract}
Thermodynamic integration (TI) offers a rigorous method for estimating free-energy differences by integrating over a sequence of interpolating conformational ensembles. However, TI calculations are computationally expensive and typically limited to coupling a small number of degrees of freedom due to the need to sample numerous intermediate ensembles with sufficient conformational-space overlap. In this work, we propose to perform TI along an alchemical pathway represented by a trainable neural network, which we term Neural TI. Critically, we parametrize a time-dependent Hamiltonian interpolating between the interacting and non-interacting systems, and optimize its gradient using a score matching objective. The ability of the resulting energy-based diffusion model to sample all intermediate ensembles allows us to perform TI from a single reference calculation. We apply our method to Lennard-Jones fluids, where we report accurate calculations of the excess chemical potential, demonstrating that Neural TI reproduces the underlying changes in free energy without the need for simulations at interpolating Hamiltonians.
\end{abstract}

\maketitle
\def\thefootnote{*}\footnotetext{balint.mate@unige.ch}
\def\thefootnote{\dag}\footnotetext{francois.fleuret@unige.ch}
\def\thefootnote{\ddag}\footnotetext{bereau@uni-heidelberg.de}
\begin{figure*}[!htb]
\begin{center}
\centerline{\includegraphics[width=\textwidth,trim={0 3.8cm 0 4.2cm},clip]{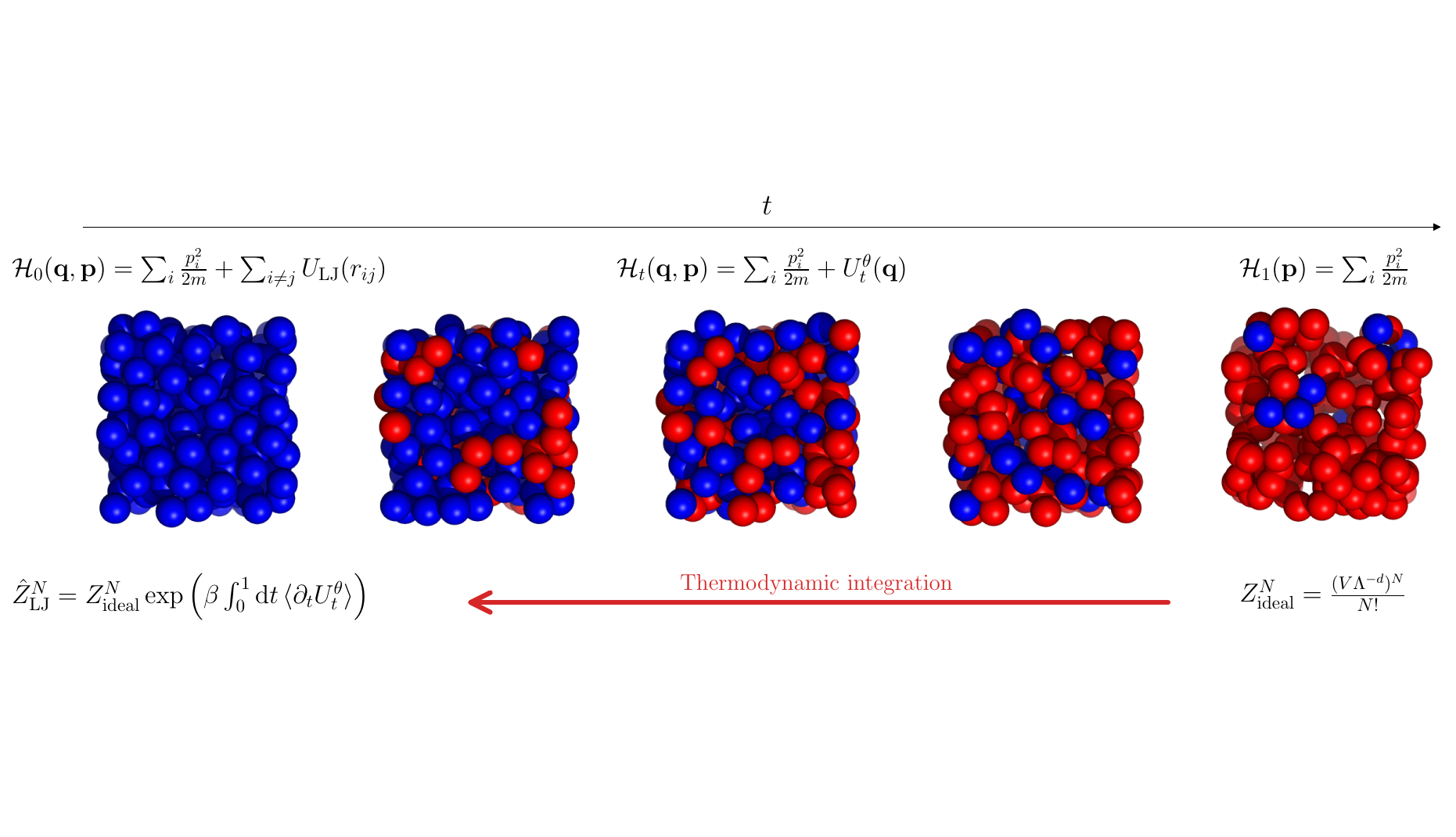}}
\vspace*{-3mm}
\caption{Schematic summary of the proposed approach. We interpolate between the target, $\mathcal H_0$, and latent, $ \mathcal H_1$, Hamiltonians with a time-dependent potential $U_t^\theta$. During sampling, the normalizing constant of the target can be estimated via thermodynamic integration. Particles whose separation from their closest neighbor is less than $0.85\sigma$ are colored red, the rest are colored blue. This color-coding illustrates that in the ideal gas (right) there are many colliding particles  and as the LJ potential is turned on (left) the particles do not overlap anymore.}
    \label{fig:summary}
\end{center}
\vskip -0.2in
\end{figure*}

\section{Introduction}
Accurate estimation of free-energy differences is pivotal in numerous scientific disciplines, including chemistry, biology, and materials science. These estimations are essential for understanding molecular interactions, reaction mechanisms, and phase transitions \cite{gao2006mechanisms, mobley2017predicting, agarwal2021free}. The main methodologies to estimate free-energy differences are rooted in statistical mechanics including free-energy perturbation and thermodynamic integration (TI). They estimate the ratio of partition functions of the two ensembles, typically sampling the two conformational ensembles via Monte Carlo or molecular dynamics simulations. Because of typically poor overlap between distributions, intermediate simulations are required to help interpolate the two end points.  The exact pathway connecting the two ensembles can be chosen freely, because the free energy is a state function. The chosen interpolation is often unphysical, resulting in a so-called \emph{alchemical transformation}. The necessity to sample intermediate Hamiltonians leads to significant computational expense \cite{chipot2007free, mey2020best}.

Machine learning---in particular generative models---is rapidly transforming how we model molecular systems \cite{sanchez2018inverse, noe2020machine, fedik2022extending}. Normalizing flows have been proposed for sampling in the context of statistical physics \cite{nicoli2020asymptotically}, lattice quantum field theory \cite{albergo2019flow}, and molecules \cite{noe2019boltzmann}. Flow-based approaches are attractive as they represent exact probability densities that can be used for unbiased estimation of observables when combined with importance sampling to correct for the mismatch between the learnt and target densities. \citet{invernizzi_skipping} have revisited parallel tempering by a learned flow map to increase phase-space overlap. Flow-based approaches have also been extended to estimate free energy differences via ML-enhanced free energy perturbation \cite{wirnsberger2020targeted} and generalized to different thermodynamical ensembles \cite{Wirnsberger_2023}. Flows, however, also present challenges: coupling flows \cite{dinh2016density} are cumbersome to work with when one needs to encode physical bias in the architecture, and continuous flows \cite{chen2018neural} suffer from the cost of divergence computations.

In this work, we instead propose to use denoising diffusion models (DDMs) \cite{sohl2015deep, ho2020denoising} to estimate free-energy differences. To this end, we  assign our data and latent distributions to the \emph{interacting} and \emph{non-interacting} Hamiltonians, respectively. 
We then train a time-dependent, interpolating potential $U_t^\theta$ between the potentials of the target, $U_0$, and the prior, $U_1$, along the finite-time interval of the DDM. We match the force exerted by the potential to the (Stein) score \cite{liu2016kernelized} of the DDM
$$
s(x,t) = \nabla_x \log \rho_t = -\beta \nabla_x U_t^\theta.
$$

This parametrization of the score, while limiting its expressivity, will easily allow us to compute time derivatives of the energy---a critical component to perform TI along the diffusion process. The TI calculation efficiently estimates the ratio of partition function of turning on the interactions of the Hamiltonian. However, because we can analytically calculate the partition function of the non-interacting system as well, our methodology yields the partition function of the target Hamiltonian. We call our method Neural TI.

We demonstrate our methodology on computer simulations of a condensed-phase, many-body system: the Lennard-Jones (LJ) liquid. The particles are confined to a box with periodic boundary conditions, leading to topological constraints on the DDM addressed further below. The latent space consists of the ideal-gas system: particles that carry kinetic energy, but do not interact. The DDM thereby interpolates between interacting and non-interacting system, whose interpolating densities correspond to the Boltzmann densities of time-dependent machine learned potential $U_t^\theta$. See Figure \ref{fig:summary} for a schematic summary of the proposed approach. We validate Neural TI on accurate calculations of the excess chemical potential. We report accurate free-energy differences of coupling all LJ interactions starting from the ideal gas, without relying on simulations at interpolating Hamiltonians.

\paragraph*{Thermodynamic integration}

Suppose now that $U_\lambda$ is a one-parameter family of potentials, and $Z_\lambda = \int\text{d}x\,e^{-\beta U_\lambda(x)}=e^{-\beta F}$ are the corresponding normalizing constants at inverse temperature $\beta = (k_\text{B}T)^{-1}$. Given samples from all $\rho_\lambda=\text{e}^{-\beta U_\lambda(x)}/Z_\lambda$  the free energy difference $\Delta F_{0\rightarrow 1}$ between $\lambda=0$ and $\lambda=1$ can be written as
{\allowdisplaybreaks
\begin{align}
    \label{eq:thermodynamic_int}
    \beta \Delta F_{0\rightarrow 1}&=
    \log Z_0-\log Z_1 \\&=
    -\int_0^1 \text{d}\lambda \, \partial_\lambda \log Z_\lambda \\
     &=  -\int_0^1 \text{d}\lambda \, \frac{1}{Z_\lambda}\partial_\lambda Z_\lambda \\
     &=  -\int_0^1 \text{d}\lambda \, \frac{1}{Z_\lambda}\partial_\lambda\left(\int \text{d}x \, \text{e}^{-\beta U_\lambda(x)}\right) \\
     &=  \beta\int_0^1 \text{d}\lambda \, \frac{1}{Z_\lambda}\left(\int \text{d}x \, \text{e}^{-\beta U_\lambda(x)} \partial_\lambda U_\lambda(x)\right) \\
     &=  \beta \int_0^1 \text{d}\lambda \, \left\langle  \partial_\lambda U_\lambda \right\rangle_\lambda, \label{eq_TI}
\end{align}
}where $\langle  \partial_\lambda U_\lambda\rangle_\lambda$ denotes the expected value of $\partial_\lambda U_\lambda$ under the density $\rho_\lambda = Z_\lambda^{-1}\text{e}^{-\beta U_\lambda}$. Practically, the small phase-space overlap between ensembles requires an interpolation of many intermediate Hamiltonians parametrized by the coupling variable $\lambda$ \cite{mey2020best}. Here instead we will use a DDM 
to learn the alchemical pathway, i.e., we replace $\lambda$ by the diffusion model's time variable, $t$. Diffusion models allow us to sample the equilibrium distribution of each intermediate state, meaning that our approach does \emph{not} coincide with non-equilibrium variants, such as slow- or fast-growth TI \cite{hummer2001fast}. The generative model's ability to sample all intermediate ensembles will do away with the need for intermediate reference simulations, and provide an accurate estimate of much larger free-energy differences than reported so far \cite{straatsma1988free}. In the machine-learning community, \citet{masrani2019thermodynamic} used TI for tightening the evidence lower bound (ELBO).

\paragraph*{Denoising Diffusion models}
Diffusion models \cite{sohl2015deep, ho2020denoising}  are a class of generative models defined by a pair (forward and reverse) of stochastic processes. In the continuous-time formulation \cite{Song2020ScoreBasedGM}, the forward process is given by the stochastic differential equation (SDE)
\begin{equation}
    \label{eq:diff_forward}
    dX = f_t X \text{d}t^{\rightarrow} + g_t \text{d}W_{t^{\rightarrow}},
\end{equation}
where $\text{d}W_{t^{\rightarrow}}$ is a Wiener process and SDE starts from the initial condition $X_0 \sim \rho_0$.
Note that $\rho_t(x_t|x_0)$ is a Gaussian for all $t$ and $x_0$. In particular $f_t$ and $g_t$ are time-dependent and  are chosen such that $\rho_t(x_1|x_0)$ is close to a standard Gaussian distribution for all $x_0$. 
The noising process then corresponds to simulating Equation \ref{eq:diff_forward} from $t=0$ to $t=1$.
The same time marginals can be reproduced by a time-reversed SDE \cite{ANDERSON1982313} from $t=1$ to $t=0$,
\begin{equation}
    \label{eq:diff_reverse}
    dY =  \left[f_t Y + g_t^2\nabla_x \log \rho_t\right] \text{d}t^{\leftarrow}+ g_t \text{d}W_t^{\leftarrow},
\end{equation}
with initial condition $Y_0 \sim \mathcal N(0,\mathbb{I})$.
Given the score $s(x,t)=\nabla_x \log \rho_t(x)$, one could use the reverse Equation \ref{eq:diff_reverse} to map Gaussian samples to the initial target density $\rho_0$.

Physically speaking,  both the forward and reverse processes correspond to an overdamped Langevin dynamics driven by the time-dependent potentials $\tfrac{1}{2}f_t||x||^2$ and $\tfrac{-1}{2}f_t||y||^2- g_t^2 \log\rho_t$, respectively. We can thus interpret the score, $\nabla \log\rho_t$, as a force induced by the potential $(-\log\rho_t)$ \cite{arts2023two}. 

\paragraph*{Score estimation on $\mathbb{R}^d$}  
To estimate the score $s(x,t) = \nabla_x \log \rho_t$, one integrates over the conditional scores
\begin{align}
    \nabla \log \rho_t(x) &= \mathbb{E}_{x_0\sim \rho_0, x_t \sim p(x_t|x_0)} \left[\nabla \log \rho_t(x_t|x_0)\right]. 
\end{align}
Denoting the mean and variance of $\log \rho(x_t|x_0)$ by $\gamma_tx_0$ and $\sigma^2_t$, we can rewrite the integrand as
\begin{align}
    \nabla \log \rho_t(x_t|x_0) &=-\frac{x_t-\gamma_t x_0}{\sigma_t^2} 
    =-\frac{\epsilon}{\sigma_t}.
\end{align}
Instead of learning the score directly, it is customary to train a neural network $\epsilon_\theta(x,t)$ to predict the  noise term $\epsilon= \frac{1}{\sigma_t}(x_t-\gamma_t x_0) $ instead. The score is then recovered as  $s_\theta(x,t) = -\epsilon_\theta(x,t)/\sigma_t$.

\paragraph*{TI with diffusion models} 
\citet{salimans2021should} raise the point that although  the score $s(x,t)$ is the gradient of the log-density $\log \rho_t(x)$, it is usually modelled with a free-form neural network that does not necessarily learn a conservative vector field. Our proposal is to approximate the score as the force of a time-dependent, parametric potential $U^\theta_t$,
$$
s(x,t) = \nabla_x \log \rho_t = -\beta \nabla_x U_t^\theta.
$$
This in turn means that $ U^\theta_t$ itself serves as an approximation of the negative log-likelihood up to an additive constant. Since $U^\theta_t $ is a neural network, its time-derivative can be computed with automatic differentiation and the ensemble average $\langle \partial_t U^\theta_t\rangle_t $ can be estimated from samples either from the forward or from the learned reverse process.

\paragraph*{Diffusion models on tori}
\label{sec:toroidal_diff}
In what follows we will be interested in particles living in a $d$-dimensional box with periodic boundary conditions. To work with such systems the usual framework of diffusion models needs to be slightly adjusted to accommodate for the different topology of the configurational space. Topologically speaking, the position of a single particle is specified by a point on a hypertorus, $\mathbb T^{N}$, of dimension $d$,  and the configurational space of $N$ particles is then a $dN$-dimensional hypertorus, $\mathbb T^{dN}$. 

Although there are works generalizing diffusion models to non-Euclidean geometries \cite{de2022riemannian,huang2022riemannian}, for our case it is sufficient to derive a model for the simplest manifold with non-trivial topology, the circle $\mathbb S^1 = \mathbb R/\mathbb{Z}$. To do this, we set the forward process to be an unbiased random walk on $\mathbb S^1$ converging to the uniform distribution. Explicitly, the forward and backward processes take the following form,
\begin{align}
    dX &=  g_t \text{d}W_{t^{\rightarrow}}\qquad\qquad\qquad\qquad\quad\, X_0 \sim \rho_0 \\
    dY &=  g_t^2\nabla \log \rho_t \text{d}t^{\leftarrow}+ g_t \text{d}W_{t^{\leftarrow}} \qquad Y_0 \sim U(\mathbb S^1).
\end{align}

\paragraph*{Score estimation on $\mathbb S^1$} 
Note that on the circle the time-marginals $\rho_t(x_t|x_0)$ are  wrapped Gaussians,
\begin{align}
    \label{eq:wrapped_gaussian}
    \rho_t(x_t|x_0) = \sum_{k\in \mathbb Z} \mathcal N(x_t + {} k;x_0,\sigma_t^2). 
\end{align}
\citet{jing2022torsional} compute the score of the wrapped Gaussian in Equation \ref{eq:wrapped_gaussian})by truncating both the numerator and denominator of the logarithmic derivative $\nabla \log \rho_t(x_t|x_0) =\frac{\nabla \rho_t(x_t|x_0)}{\rho_t(x_t|x_0)}$. Alternatively, one could write the score at $x_t$ of the wrapped Gaussian as a weighted average of the scores of the unwrapped Gaussian over the set of points that are mapped to  $x_t$ under the function $x \mapsto x \,\mathrm{mod}\, 1$,
\begin{align}
    \nabla \log \rho_t(x_t|x_0) 
    &= \frac{\nabla \rho_t(x_t|x_0)}{\rho_t(x_t|x_0)}  \\
    &= \frac{\sum_{k\in \mathbb Z}  \nabla \mathcal N_k }{\sum_{k\in \mathbb Z} \mathcal  N_k} \\
    &= \frac{\sum_{k\in \mathbb Z} \mathcal N_k \frac{\nabla \mathcal N_k }{\mathcal N_k}}{ \sum_{k\in \mathbb Z}\mathcal  N_k} \\
    &= \frac{\sum_{k\in \mathbb Z} \mathcal N_k \nabla \log \mathcal N_k}{ \sum_{k\in \mathbb Z}\mathcal  N_k},
\end{align}
where $\mathcal N_k =\mathcal N(x_t + {} k;x_0,\sigma_t^2)$ are the evaluations of the unwrapped Gaussian on the fiber and $\sigma_t^2=\int_0^t \text{d}\tau \, g_\tau^2$. This in particular means that the noise prediction model $\epsilon_\theta$ can be trained by sampling $x_t = [(x_0 + \sigma_t \epsilon) \mod 1]$ where $\epsilon \sim \mathcal N(0,1)$ and taking a gradient step on $||\epsilon_\theta(x_t,t)-\epsilon||^2$.

\section{Statistical ensembles}

For the rest of the paper we exclusively consider systems of indistinguishable particles confined to a $d$-dimensional box of  volume $V$ with periodic boundary conditions. 

\paragraph*{Canonical ensemble ($NVT$)} 
If  we assume that a system has a fixed number of particles $N$ and is in thermal equilibrium with a reservoir at a fixed inverse temperature $\beta= (k_\text{B}T)^{-1}$, then the likelihood of a particular microstate $(\mathbf q,\mathbf p)=(q_1,...q_N, p_1,...,p_N)$ is 
\begin{equation}
    \rho(\mathbf q,\mathbf p) =\frac{1}{Z_N} \frac{1}{h^{dN}N!}\text{e}^{-\beta \mathcal{H}(\mathbf q,\mathbf p)},
\end{equation}
where $h$ is Planck's constant, $d$ is the dimensionality of the system, $\mathcal{H}$ is the Hamiltonian and $Z_N = \int \frac{\text{d} \mathbf p \text{d}\mathbf q}{h^{dN}N!}\text{e}^{-\beta \mathcal{H}(\mathbf q,\mathbf p)}$ is the canonical partition function, i.e., the normalizing constant of the density $\frac{\text{e}^{-\beta \mathcal{H}(\mathbf q,\mathbf p)}}{h^{dN}N!}$.

\paragraph*{Ideal and interacting gases}
In the ideal gas particles do  not interact with each other and the system is described by a Hamiltonian only containing a kinetic term $\mathcal{H}_{\text{ideal}}(\mathbf p) = \sum_i \frac{p_i^2}{2m}$, where $m=m_1 = ...=m_N$ denotes the mass of all particles.
In this case $Z_N$ can be analytically computed
\begin{align}
    Z_N^{\text{ideal}}=\int \frac{\text{d} \mathbf p}{h^{dN}N!}\text{e}^{-\beta \mathcal{H}_{\text{ideal}}(\mathbf p)}=\frac{(V\Lambda^{-d})^ N}{N!}, \label{eq:Z_ideal_gas}
\end{align}
where $V$ is the volume of the box and $\Lambda = h / \sqrt{2 \pi m \beta^{-1}}$ is the thermal wavelength. Since particle positions in the ideal gas are independently and uniformly distributed over the box, the latent space of the toroidal diffusion model 
describes the positions $\mathbf q$ of an ideal gas in the canonical ensemble.

More generally, a many-body system will also consist of interactions, as described by a potential $U(\mathbf q)$,
\begin{equation}
     \mathcal{H}(\mathbf q,\mathbf p) = \sum_i \frac{p_i^2}{2m} + U(\mathbf q).
\end{equation}
The separation between the positions and momenta offers a factorization of the partition function
\begin{equation}
    Z_N = Z_N^{\text{ideal}} \int \text{d} \mathbf q \,\text{e}^{-\beta U(\mathbf q)}.
\end{equation}
\paragraph*{Estimating the partition function}
Suppose now that we have trained a DDM on the positions $\mathbf q$ with its score parametrized as the force of a time-dependent potential, $s_\theta(\mathbf q,t)=-\beta\nabla U^\theta_t(\mathbf q)$, between the target and non-interacting Hamiltonians, i.e., $U^\theta_0(\mathbf q)=U(\mathbf q)$ and $U^\theta_1(\mathbf q)  = U_{\text{ideal}}(\mathbf q)\equiv 0$. As the toroidal DDM maps configurations of the ideal gas to the interacting system, the setup is adequate to perform TI over the coupling of interactions. In addition, because $Z_N^\text{ideal}$ is known analytically, we obtain an estimate of the partition function of the full, target Hamiltonian
\begin{align}
\label{eq:Z_N_estimate}
    \hat Z_N = Z_N^{\text{ideal}} \exp\left(\beta\int_0^1 \text{d}t\, \langle \partial_t U_t^\theta\rangle_{t}\right).
\end{align}

\paragraph*{Grand Canonical ensemble ($\mu VT$) }
Let us now assume that a system is both in thermal and chemical equilibrium with the reservoir at fixed temperature $T$ and chemical potential $\mu$. The likelihood of a microstate $(N,\mathbf q,\mathbf p)=(N,q_1,...q_N, p_1,...,p_N)$ is 
\begin{equation}
    \label{eq:grand_canonical_ensemble}
    \rho(N,\mathbf q,\mathbf p) =\frac{1}{\mathcal Z} \frac{1}{h^{dN}N!}\text{e}^{\beta (\mu N-\mathcal{H}(\mathbf q,\mathbf p))},
\end{equation}
where $\mathcal Z = \sum_N\int \frac{\text{d} \mathbf q\text{d} \mathbf p}{h^{dN}N!}\text{e}^{\beta (\mu N-\mathcal{H}(\mathbf q,\mathbf p))}$
is the grand canonical partition function. The grand canonical partition function is a weighted sum of the canonical partition functions according to the chemical potential,
\begin{equation}
   \mathcal Z(\mu)  = \sum_N \text{e}^{\beta\mu N } Z_N.
\end{equation}

\paragraph*{The excess chemical potential}
The choice of a  Hamiltonian $\mathcal H$ and chemical potential $\mu$ defines a marginal distribution, $p(N)$, of the number of particles present in the system 
\begin{equation}
    \label{eq:pN_grand_canonical}
    p(N) = \frac{1}{\mathcal Z}\text{e}^{\beta\mu N } Z_N.
\end{equation}
In the case of the ideal gas, $p(N)=\frac{1}{\mathcal Z}\frac{(\text{e}^{\beta\mu}V\Lambda^{-d})^ N}{N!}$ is Poisson distributed with expected value 
\begin{equation}
    \langle N \rangle (\mathcal H^{\text{ideal}},\mu) = \text{e}^{\beta\mu}V\Lambda^{-d}.
\end{equation}

For interacting systems, the expected number of particles $\langle N \rangle(\mathcal H,\mu)$ can be used  to decompose the chemical potential as $\mu = \mu_{\text{ideal}}+\mu_{\text{ex}}$, where $\mu_{\text{ideal}}$ is the chemical potential of the ideal gas of the same density
\begin{equation}
    \label{eq:mu_ideal}
    \mu_{\text{ideal}}=\beta^{-1}\log\frac{\langle N \rangle \Lambda^d}{V}.
\end{equation}

Intuitively, the excess chemical potential $\mu_{\text{ex}} = \mu-\mu_{\text{ideal}}$ measures the deviation in chemical potential due to the interaction term of the Hamiltonian at fixed density.
The straightforward way of estimating the excess chemical potential  of an interacting system is to perform a grand-canonical Monte Carlo (GCMC) simulation on the distribution in Equation \ref{eq:grand_canonical_ensemble} at a prescribed value of $\mu$. Grand-canonical sampling of the number of particles allows us to estimate the empirical mean $\langle N\rangle_{\text{GCMC}}$, from which we extract $\mu_\text{ideal}$ via Equation \ref{eq:mu_ideal}. From $\mu$ and $\mu_\text{ideal}$, one can simply compute the excess chemical potential, $\mu_\text{ex} = \mu - \mu_\text{ideal}$.

In this work we estimate the excess chemical potential from several canonical simulations. We train a generative model on several canonical ensembles and then estimate the canonical partition function $Z_N$ for each $N$. The collection $\{Z_N\}$ is used to compute the distribution of the number of particles, $p(N)$, see Equation \ref{eq:pN_grand_canonical}, for a predefined choice of the chemical potential, $\mu$. In combination with Equation \ref{eq:mu_ideal}, we compute the excess chemical potential, $\mu_\text{ex}$. 

\section{Application: Lennard-Jones fluid}
\label{sec:exeperiments}

We now consider a many-body condensed-phase system: a collection of three-dimensional particles confined in a box, interacting via a pairwise Lennard-Jones (LJ) potential
\begin{align}
    U_\text{LJ}(\mathbf q)= \sum_{i\neq j}4\varepsilon\left[\left(\frac{\sigma}{r_{ij}}\right)^{12}-\left(\frac{\sigma}{r_{ij}}\right)^{6}\right],
\end{align}
where $r_{ij}$ represents the inter-particle distance between particles $i$ and $j$.

To demonstrate the accuracy of our DDM-based Neural TI scheme, we perform the following
\begin{enumerate}
    \item Generate training configurations via canonical Monte Carlo simulations at different particle numbers $N_1,\dots,N_i$;
    \item Train a \emph{single} diffusion model on the positions $\mathbf q$ of the training data, making the model transferable between different $N_i$. This transferability property is crucial since we would like to estimate $Z_N$ also at those values of $N$ that did not belong to the training set;
    \item Estimate the partition function, $Z_N$, by generating configurations at a given $N$ from the DDM and computing $\hat Z_N$ from TI, see Equation \ref{eq:Z_N_estimate};
    \item Make grand-canonical estimates from a collection of canonical $\hat Z_N$ across values of $N$, $\{\hat Z_{N_1}, \hat Z_{N_2}, \dots \}$, and compare against reference GCMC simulations.
\end{enumerate}

\paragraph*{Parametrization of the score}
To ensure that the boundary conditions at $t\in \{0,1\}$ of the energy interpolation are met, we follow the interpolation proposed by \citet{mate2023learning} and parametrize the score as
\begin{align}
    \label{eq:score_param}
    s(\mathbf q,t) = -\beta \nabla\left[t(1-t)U_t^\theta(\mathbf q) + (1-t) U_t^\text{LJ}(\mathbf q)\right],
\end{align}
where $U_t^\text{LJ}$ is a soft-core LJ potential \cite{BEUTLER1994529} with a time-dependent softening parameter
\begin{align}
    \label{eq:soft_LJ}
    U_t^\text{LJ}(\mathbf q)= \sum_{i\neq j}4\varepsilon \left[\left(\frac{\sigma^2}{t\sigma^2+ r_{ij}^2}\right)^{6}-\left(\frac{\sigma^2}{t\sigma^2+r_{ij}^2}\right)^{3}\right].
\end{align}

\begin{figure}[h]
\begin{center}
\centerline{\includegraphics[width=.9\columnwidth,trim={0 .3cm 0 .2cm},clip]{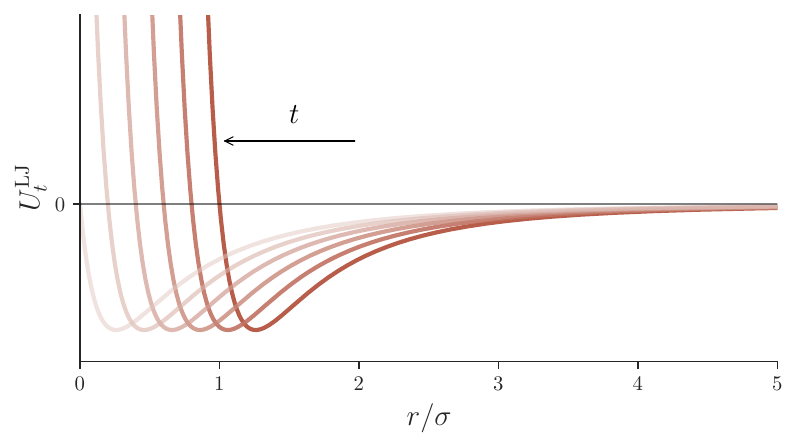}}
\vspace*{-3mm}
\caption{The soft-core LJ potential $U_t^\text{LJ}$ in Equation \ref{eq:soft_LJ} for various values of $t \in [0,1]$. Note that for larger values of $t$, particles can get closer to each other without experiencing strong repulsive forces. This is necessary since in the diffusion process it is inevitable that particles get close to each other as $t$ increases.}
    \label{fig:soft_LJ}
\end{center}
\vskip -0.2in
\end{figure}

This soft-core potential ensures that the numerically unstable region of the LJ potential is not evaluated on noisy samples but is slowly introduced as more and more noise is removed from the samples (Figure \ref{fig:soft_LJ}). The only trainable component, $U^\theta_t$ can then be parametrized using ideas from the machine learning force field literature \cite{schutt2017schnet,gasteiger2019directional,batatia2022mace,batzner20223}. The parametrization in  Equation (\ref{eq:score_param}) means that the neural network learns the deviation from the pathway typically used in the standard approach to TI.

\paragraph*{Results} 
We set the dimension of the system to $d=3$, the volume of the box to $V=216$, the inverse temperature to $\beta=1$, the mass of the particles to
$m = 1$ and the parameters of the LJ potential to $\varepsilon = 0.8,\sigma=1$. We set the value of Planck's constant to $10^{-4}$, but stress that  the observables we report do not depend on $h$, its numerical value was chosen out of convenience. We train a \emph{single} DDM on samples from canonical Monte Carlo simulations at particle numbers $N\in \{40,80,120,160,200\}$ (i.e. densities $\rho=N/V \in \{0.19,0.37,0.56,0.74,0.93\}$). We refer the reader to the Appendix
\ref{appendix:details} 
for details on the Monte Carlo simulations and on the architecture.

\begin{figure}[!htb]
\vskip -0.1in
\begin{center}
\centerline{\includegraphics[width=\columnwidth,trim={0 1.4cm 0 .2cm},clip]{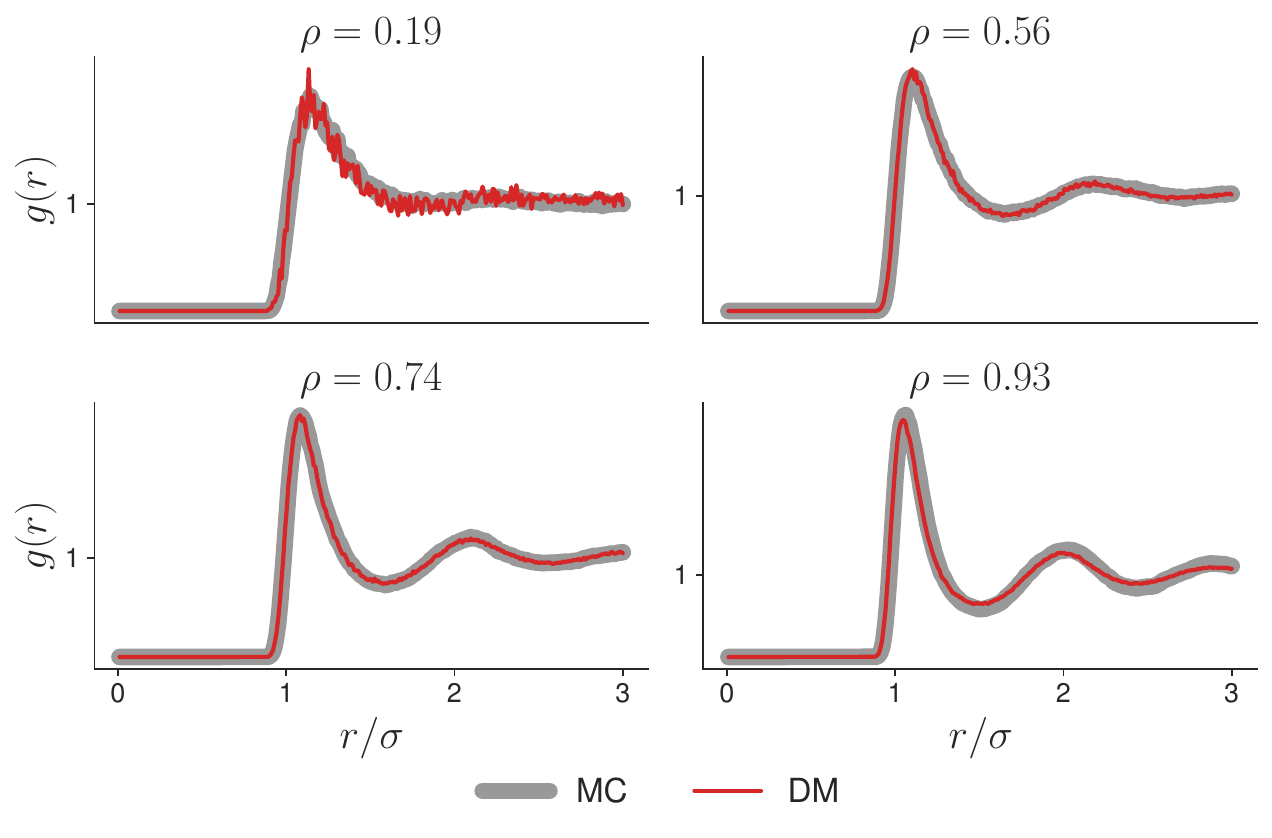}}
\vspace*{-3mm}
\caption{ Radial distribution functions as predicted by Monte Carlo simulations (gray) and a diffusion model (red) trained on densities $\rho \in \{0.19,0.37,0.56,0.74,0.93\}$. Note that the model reconstructs $g(r)$ across the the gas-liquid phase transition.}
\label{fig:LJ3D_gr}
\end{center}
\vskip -0.2in
\end{figure}

To evaluate the generative performance of these models we compare the radial distribution function (RDF), $g(r)$, between Monte Carlo samples and samples from the trained DDM (Figure \ref{fig:LJ3D_gr}). We find accurate reconstructions across a broad range of densities, $\rho = 0.19$ to $\rho = 0.93$. The shapes of the RDF clearly indicate a transition from gas to liquid as we increase particle density. Note that reconstructing observables from generated samples hints at a good fit between the learnt and target densities. It is a necessary but not sufficient condition for the accuracy of TI along the model.

\begin{figure}[!htb]
\vskip -0.1in
\begin{center}
\centerline{\includegraphics[width=\columnwidth,trim={0 1.35cm 0 .2cm},clip]{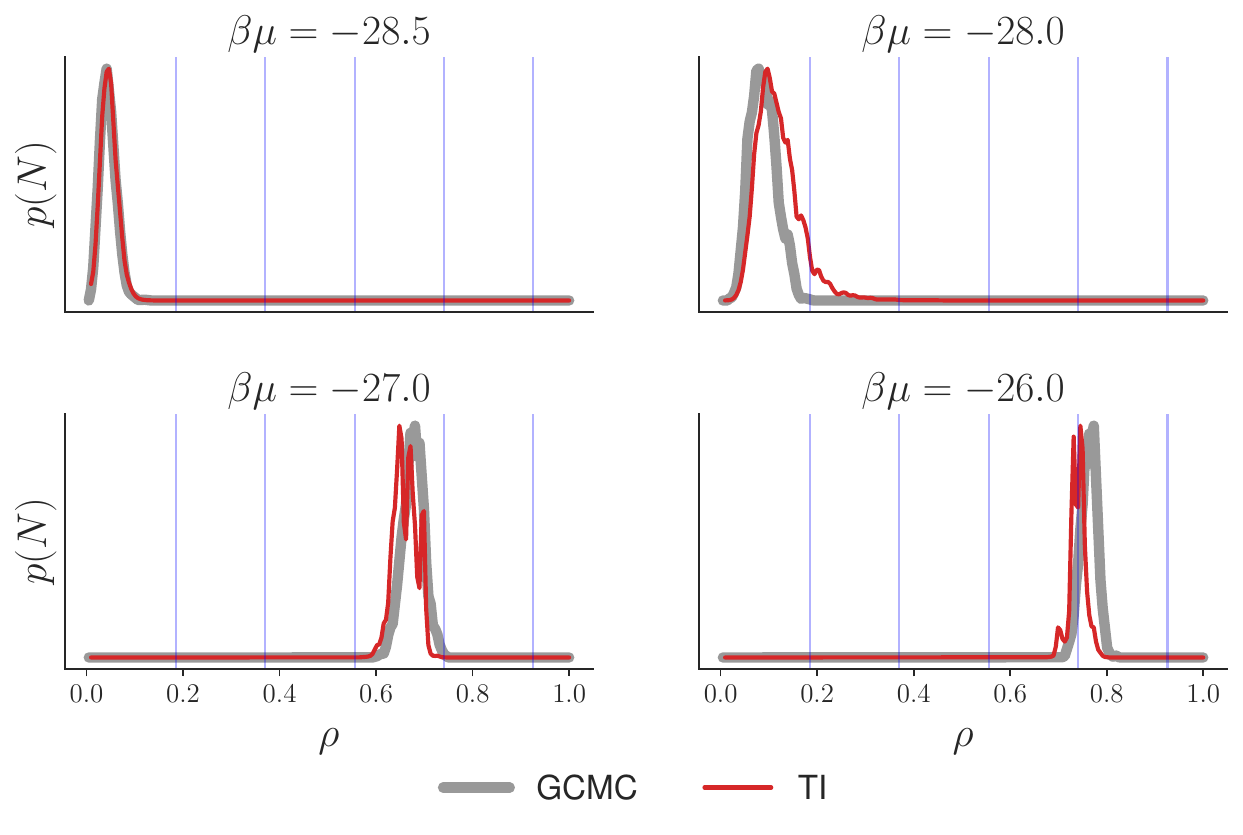}}
\vspace*{-3mm}
\caption{ Distribution of the number of particles in the grand canonical ensemble at different chemical potentials from GCMC simulations (gray) and estimated by thermodynamic integration with a trained diffusion model (red). The vertical blue lines denote the canonical ensembles that the diffusion model was trained on.}
\label{fig:LJ3D_pn}
\end{center}
\vskip -0.2in
\end{figure}

The accuracy of the Neural TI  estimated free energy differences is assessed first in Figure \ref{fig:LJ3D_pn}.  The different panels monitor estimates of the particle-number distribution, $p(N)$, and show significant changes in the distribution as we change the chemical potential. They highlight the transferability of our methodology across the phase transition.

\begin{figure}[!htb]
\vskip -0.1in
\begin{center}
\centerline{\includegraphics[width=\columnwidth,trim={0 0 0 0},clip]{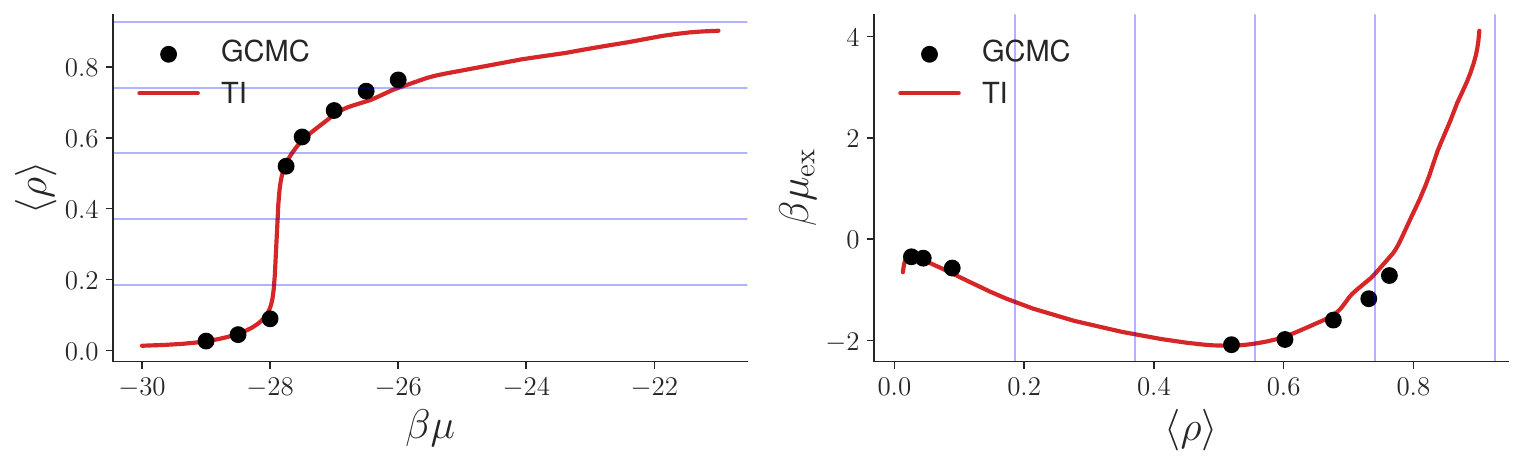}}
\vspace*{-3mm}
\caption{ Expected density as a function of the chemical potential (left) and estimates of $\mu_\text{ex}$ as a function of the expected density (right). The left plot suggests a that a gas-liquid phase transition takes place at  $\beta\mu \approx -28$. The  blue lines denote the canonical ensembles that the diffusion model was trained on.}
\label{fig:LJ3D_mu_pn}
\end{center}
\vskip -0.2in
\end{figure}

Further, we also evaluate the functional relationship between $\rho$ and $\mu$ in Figure \ref{fig:LJ3D_mu_pn}. The left subfigure shows the average density as a function of the chemical potential, as calculated by both reference GCMC and our proposed TI methodology. Though the reference calculations are run in the grand-canonical ensemble, we make use of TI from a set of \emph{canonical} simulations to estimate $\mu$. The TI calculations estimate the partition functions, $\hat Z_N$, at different values of $N$. From those, we compute the particle-number distribution via Equation \ref{eq:pN_grand_canonical}, extract the mean, and deduce the average particle density, $\langle \rho \rangle$. We find excellent agreement across the gas--liquid phase transition---visually illustrated by the jump in density. The right subfigure shows the excess chemical potential as a function of particle density. Our methodology allows us to smoothly interpolate across the phase transition. At high density, the slight underestimation of the particle-number distribution  leads a small discrepancy in the excess chemical potential. The main workhorse-method to compute excess chemical potentials from canonical simulations, the Widom particle-insertion method, shows significant difficulties in this regime due to its perturbative nature \cite{widom1963some, frenkel2023understanding}.

\begin{figure}[!h]
\begin{center}
\centerline{\includegraphics[width=\columnwidth,trim={0 0 0 0},clip]{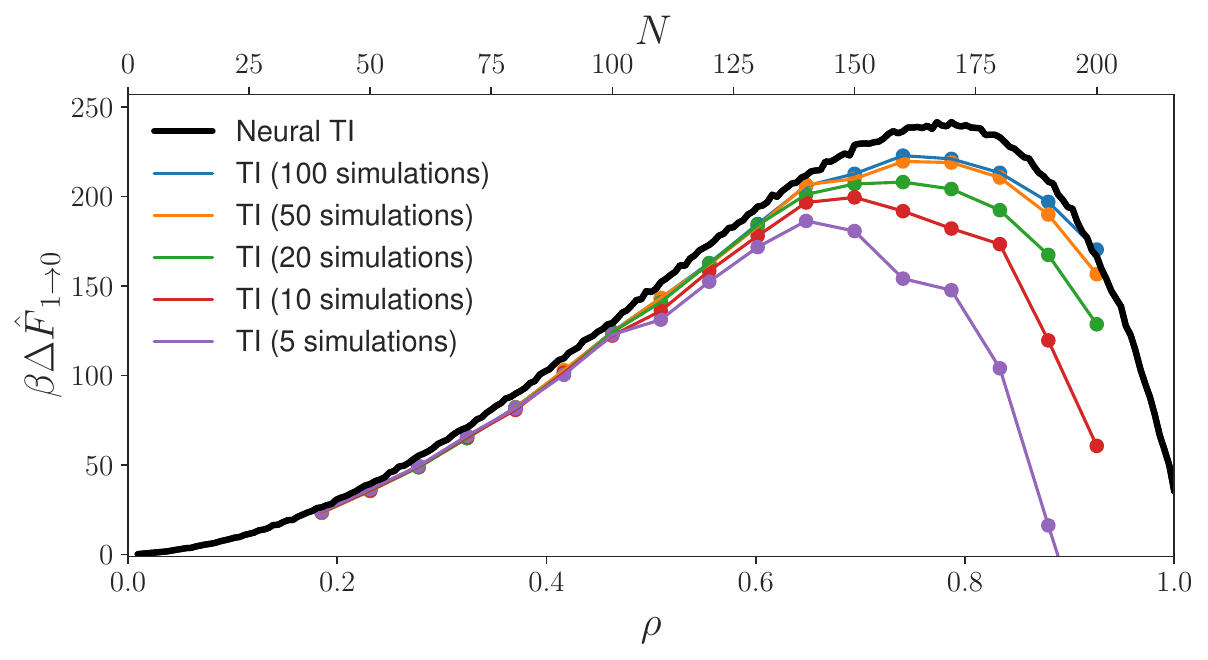}}
\vspace*{-3mm}
\caption{Estimated free energy of turning on the LJ interactions in the canonical ensemble at different densities (number of particles at the top). The subscript $1\rightarrow 0$ corresponds to an interpolation from the non-interacting latent space to the target Hamiltonian, i.e., turning on interactions in the system. Estimate by Neural TI shown in thick black. For comparison, we also display various standard TI estimates with increasing numbers of reference simulations (various colors).}
\label{fig:rho_FE}
\end{center}
\vskip -0.2in
\end{figure}

This experiment also offers insight in the equilibrium properties of liquids by directly reporting the free energy of coupling all LJ interactions starting from the ideal gas. Figure \ref{fig:rho_FE} shows the coupling free energy starting from the ideal gas to the target Hamiltonian of the LJ particle configurations at different densities. On the left, densities close to 0  lead to virtually no free energy due to the ideal-gas limit. As density increases close to $\rho\approx 0.75$, we observe a maximum---interestingly quite a bit higher in density compared to where the excess chemical potential attains its minimum, i.e., $\langle \rho \rangle \approx 0.55$. Instead, we hypothesize that the peak occurs where significant particle-overlap starts severely hampering the configurational space available---this is exactly the regime where perturbative methods, e.g., Widom insertion, become challenging. At a density of $1.0$, the available configurational integral reduces significantly---as can be seen by the drop to low $\Delta \hat F$ values. 

For comparison, we  perform the same estimate using standard TI with $5,10,20,50$ and $100$ evenly spaced intermediate simulations. The convergence of the standard TI curves with the number of intermediate simulations validates Neural TI as a free-energy estimator. At high densities the slight discrepancy between Neural TI and standard TI with many simulations is likely both due to convergence issues of the reference simulations and limited expressivity of our time-dependent ML force-field architecture.

We finally point at the sheer amplitude of the estimated free-energy differences: up to $200~k_\text{B}T$ in free energies are reported. Traditional TI methodologies require tens of interpolating Hamiltonians. Neural TI's ability to estimate free energy differences from a single reference simulation  demonstrates the appeal to \emph{learn} the alchemical pathway. While traditional alchemical-transformation methods focus on minimizing the change in Hamiltonian to a handful of degrees of freedom, here we report free-energy differences of coupling all LJ interactions to a box of ideal-gas particles.

\section{Discussion and Limitations}
Possibly the simplest way to summarize the idea behind this work is to state that energy-based parametrization combined with TI extracts a likelihood estimate from diffusion models at a relatively low cost. Given the convergence of continuous-time normalizing flows and diffusion models \cite{albergo2019flow,lipmanflow}, it is worth comparing our approach to Boltzmann generators \cite{noe2019boltzmann}. The key difference is that, unlike our approach, flows provide an exact likelihood of the model density that can later be used for importance sampling. This exact likelihood, however, requires the divergence of a vector field, which is considerably more expensive to compute than the temporal derivative of a scalar energy. Diffusion models equipped with neural TI thus offer an appealing alternative to flow-based sampling of molecular systems: an approximate likelihood at low cost. 

We can also think of the proposed approach as thermodynamic integration with an arbitrary number of intermediate $\lambda$-slices, inversely proportional to the step size of the integrator of the reverse dynamics. As with standard TI, using more steps increases accuracy but also computational cost. Our approach, however, bypasses the burden of setting up intermediate simulations, as the reverse dynamics can generate i.i.d. samples from all intermediate distributions once the model is trained. The computational tradeoff is thus between performing the intermediate simulations and training a diffusion model. In our experiment on the LJ system, neither neural TI nor standard TI was optimized for speed, and required comparable times to perform on the same hardware. We leave a detailed analysis of the computational aspects and scaling properties of the method for future work.

Looking towards chemically relevant applications, molecular systems contain a range of important physico-chemical interactions.  Here we demonstrated the applicability of Neural TI to LJ interactions---a prominent model for repulsion and dispersion interactions.  Extensions of the method to intramolecular interactions, as well as long-range Coulomb electrostatics, would require an adaptation of the energy function in Equation \ref{eq:score_param}.

\paragraph*{Conclusion}
We present Neural TI: a generative machine-learning approach to perform thermodynamic integration in molecular systems. Our work shows that energy-based denoising diffusion models (DDMs) are particularly well suited to calculate the free-energy difference of turning on interactions in a many-body Hamiltonian. We find that by associating the latent space to the ideal-gas system, and further parametrizing the score as the derivative of an energy function, we can efficiently and accurately perform TI. Unlike conventional applications, our approach does \emph{not} require reference Monte Carlo or molecular dynamics simulations at intermediate couplings between the end points. Instead, DDMs integrate the conformational ensemble along the finite-time interval of the diffusion process. We demonstrate  the accurate estimation of the free-energy difference of coupling the LJ interactions from a box of ideal-gas particles.

We demonstrate the applicability of Neural TI on Lennard-Jones fluids. We show our DDM model to transfer across densities around the gas--liquid transition of the system. The structural accuracy of the configurations is illustrated by the radial distribution functions. More importantly, we show that our TI calculations are accurate for varying numbers of particles: the excess chemical potentials and particle-number distributions result from weighted averages of TI-estimated canonical partition functions. We report free-energy differences of coupling Hamiltonians of up to $200~k_\text{B}T$ from a \emph{single} reference simulation---the fully interacting Hamiltonian alone.

\paragraph*{Acknowledgements}
We thank Fred Hamprecht and Daniel Nagel for fruitful discussions. We also thank Chin-Wei Huang for bringing the work of \citet{masrani2019thermodynamic} to the authors' attention. BM acknowledges financial support by the Swiss National Science Foundation under grant number CR - SII5 - 193716 (RODEM). TB acknowledges support by the Deutsche Forschungsgemeinschaft (DFG, German Research Foundation) under Germany's Excellence Strategy EXC 2181/1 - 390900948 (the Heidelberg STRUCTURES Excellence Cluster).

\paragraph*{Data availability}
Our implementation is available at 
\href{https://github.com/balintmate/neural-thermodynamic-integration}{https://github.com/balintmate/neural-thermodynamic-integration}.
\bibliography{biblio}


\appendix
\clearpage
\section{Experiments on a 1D Lennard-Jones system}
\label{sec:exp1d}

In this experiment we work with a $d=1$-dimensional system. We set the volume (length) of the box to $V=100$, the inverse temperature $\beta=1$, the mass of the particles to
$m = 1$, Planck's constant to $h=10^{-4}$, and the parameters of the LJ-potential to $\varepsilon = 0.8,\sigma=1$.

We train a \emph{single} diffusion model on samples from canonical Monte Carlo simulations at particle numbers $N\in \{50,70,90\}$ (i.e. densities $\rho \in \{0.5,0.7,0.9\}$). The architecture of the potential network is a SchNet-like \cite{schutt2017schnet} architecture with time-dependent RBF kernels. 

\begin{figure}[H]
\vskip -0.1in
\begin{center}
\centerline{\includegraphics[width=\columnwidth,trim={0 1.4cm 0 0},clip]{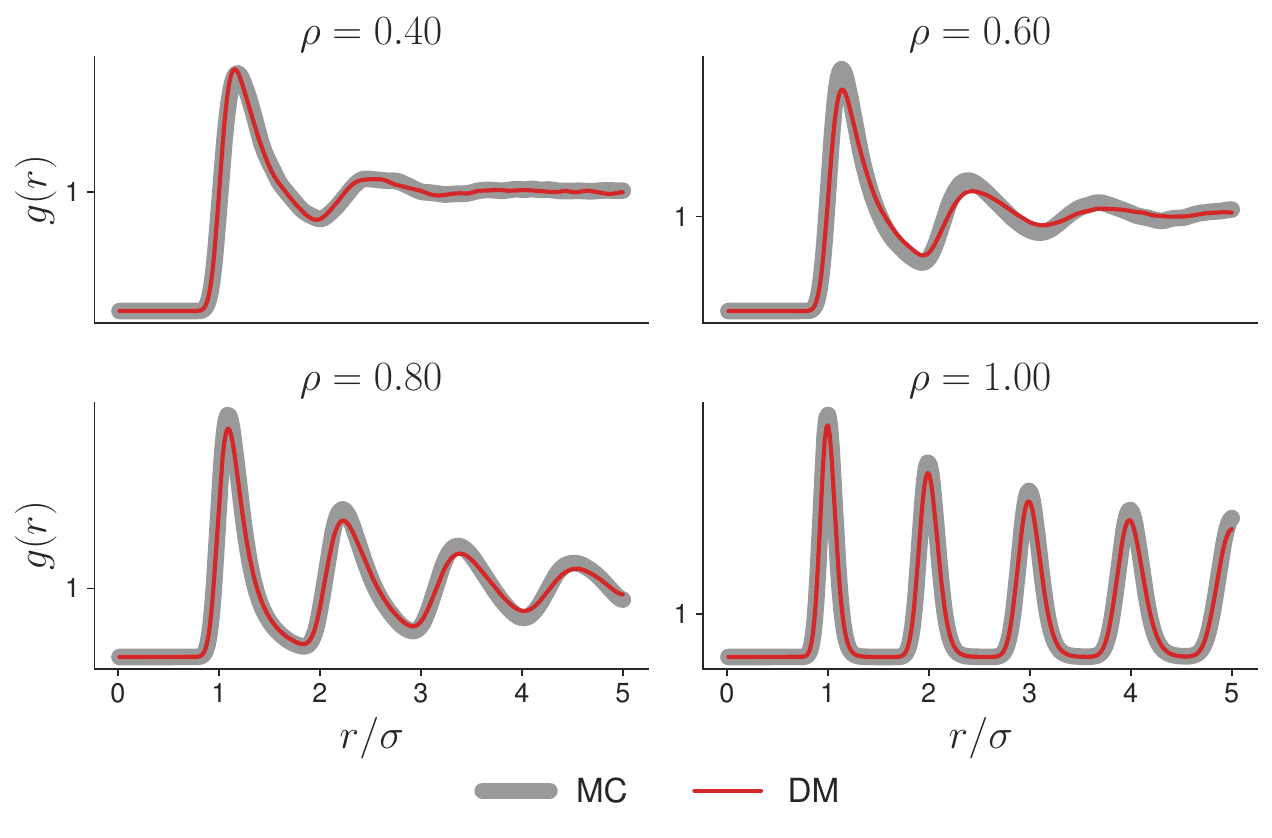}}
\vspace*{-3mm}
\caption{Radial distribution functions as predicted by Monte Carlo simulations (gray) and a diffusion model (red) trained on densities $\rho \in \{0.5,0.7,0.9\}$. Note that the model reconstructs $g(r)$ on a wide range of densities starting from $\rho=0.4$, close to the gas phase to $\rho=1.0$ close to the solid phase.}
\label{fig:LJ1D_gr}
\end{center}
\vskip -0.2in
\end{figure}

\begin{figure}[H]
\vskip -0.2in
\begin{center}
\centerline{\includegraphics[width=\columnwidth,trim={0 1.4cm 0 0},clip]{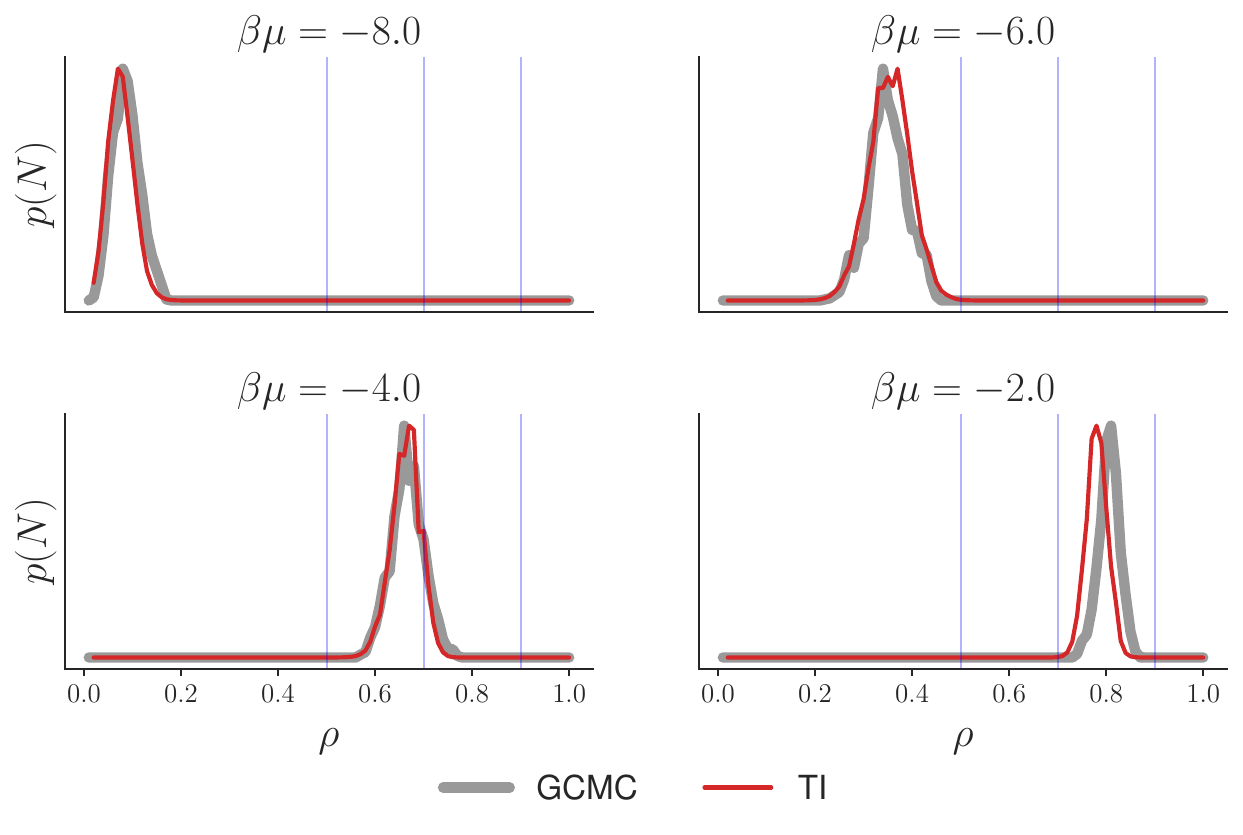}}
\vspace*{-3mm}
\caption{Distribution of the number of particles in the grand canonical ensemble at different chemical potentials as predicted by GCMC simulations (gray) and by thermodynamic integration with a trained diffusion model (red). The blue lines denote the canonical ensembles that the diffusion model was trained on.}
\label{fig:LJ1D_pn}
\end{center}
\vskip -0.2in
\end{figure}

To evaluate the generative performance of these models we compare the radial distribution function $g(r)$ between Monte Carlo samples, and samples from the trained diffusion model (Figure \ref{fig:LJ1D_gr}).  

\begin{figure}[H]
\vskip 0.2in
\begin{center}
\centerline{\includegraphics[width=\columnwidth,trim={0 0 0 0},clip]{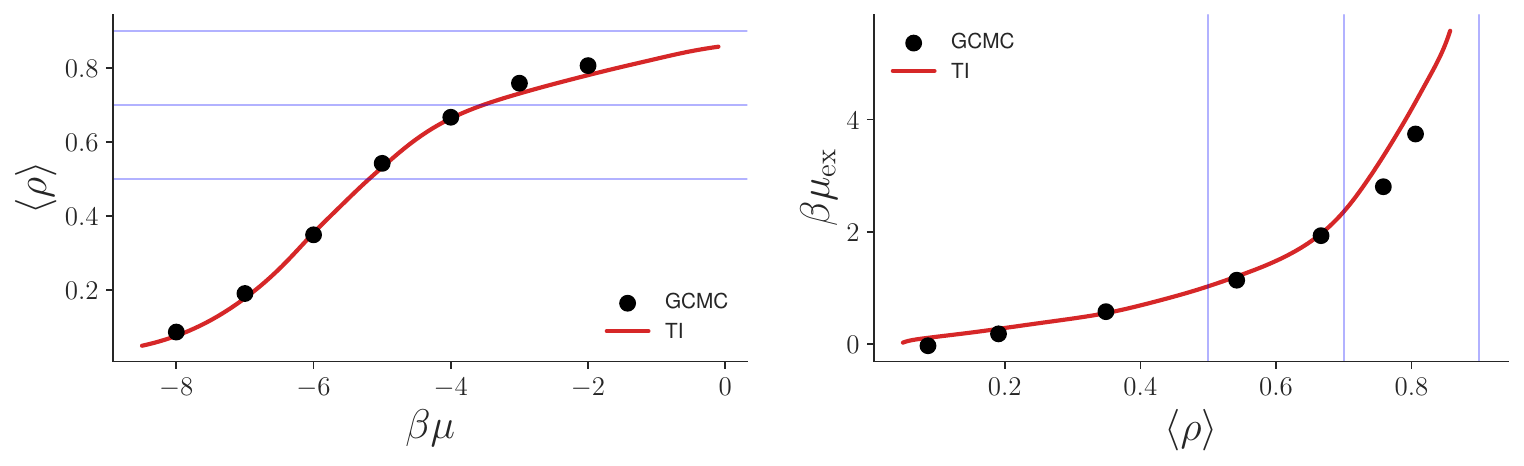}}
\vspace*{-3mm}
\caption{Expected density as a function of the chemical potential (left) and estimates of $\mu_\text{ex}$ as a function of the expected density (right).  The vertical blue lines denote the canonical ensembles that the diffusion model was trained on.}
\label{fig:LJ1D_mu_pn}
\end{center}
\vskip -0.2in
\end{figure}

The accuracy of the thermodynamic integration along the trained diffusion model is gauged by comparing its estimates of $p(N)$ (Figure \ref{fig:LJ1D_pn}) and of the relation between $\rho,\mu$ and $\mu_\text{ex}$ (Figure \ref{fig:LJ1D_mu_pn})  to grand canonical Monte Carlo simulations.

$$U_\lambda^\text{LJ}(\mathbf q)= \sum_{i\neq j}4(1-\lambda)\varepsilon \left[\left(\frac{\sigma^2}{\lambda\sigma^2+ r_{ij}^2}\right)^{6}-\left(\frac{\sigma^2}{\lambda\sigma^2+r_{ij}^2}\right)^{3}\right].$$ We collect $12000$ samples from which discard the first $2000$ as the burn-in phase.

\subsection*{Implementation details} 
We minimize a noise prediction objective $\mathbb{E}_{t\sim \mathcal U(0,1)}\mathbb{E}_{x_0\sim \rho_0}\mathbb{E}_{\epsilon\sim \mathcal N(0,\mathbf I)}||\epsilon_\theta(x_t,t)-\epsilon||^2$.
We use the Adam optimizer with its learning rate starting from $10^{-3}$ and exponentially decayed to $10^{-5}$ over 100,000 training steps. The noise schedule of the DDM is chosen to be $\sigma_t = \sigma_{min}^{1-t}\sigma_{max}^t$ with $\sigma_{min}=10^{-3}$ and $\sigma_{max}=0.5$. The value of $\sigma_{min}$ was chosen to be much smaller than the length scale of the interaction potential, whereas $\sigma_{max}$ was set to be comparable to the size of the simulation box. At sampling time, potential issues of ``lagging behind'' can appear, but are quite distinct from, say, a system driven out of equilibrium by means of a time-dependent Hamiltonian. To mitigate this, we  integrate the reverse dynamics with $1000$ time steps.

\paragraph*{ D=1} For the $1$-dimensional box 

we use a SchNet-like \cite{schutt2017schnet} architecture. To make the architecture time-dependent, we predict the parameters of the RBF kernels from the diffusion time using a small MLP. The network consists of 3 layers each of which performs a message passing step and an atom-wise update. The final readout of the energy is the sum of the features over all nodes and channels. The number of channels is 32 and the MLPs in the RBF-parameter prediction and in the atom-wise update both have two hidden layers with 64 neurons. Since the number of particles is reasonably low ($\leq 100$), we do not use  a cutoff radius, and perform message passing on the fully connected graph between the nodes.

\paragraph*{D=3} For the $3$-dimensional box 
we find that the SchNet-like architecture that worked in the one-dimensional case, can not reconstruct $g(r)$ above a density of $~0.30$.
We thus include directional information \cite{gasteiger2019directional} to the potential network. In our architecture  nodes and edges are equipped with both scalar and vectorial features. For this discussion we denote these features by $h_\text{scalar}^\text{node},\,h_\text{vec}^\text{node},h_\text{scalar}^\text{edge},\,h_\text{vec}^\text{edge}$. Our network consists of 3 layers each of which performs the following sequence of steps.
\begin{enumerate}
    \item Update $h_\text{scalar}^\text{edge}$ ($h_\text{vec}^\text{edge}$) as a linear combination of the current value of $h_\text{scalar}^\text{edge}$ ($h_\text{vec}^\text{edge}$) and the values of $h_\text{scalar}^\text{node}$ ($h_\text{vec}^\text{node}$) of the  source and target nodes.
    \item Compute $g_\text{scalar}^\text{node}$ ($g_\text{vec}^\text{node}$) by aggregating the edge features at the target nodes with time-dependent RBF weights (see previous paragraph).
    \item Update $h^\text{node}_\text{vec}$ as a linear combination of $h^\text{node}_\text{vec}$ and $g_\text{vec}^\text{node}$.
    \item Update $h^\text{node}_\text{scalar}$ as a MLP whose input is $h^\text{node}_\text{scalar}$, $g_\text{vec}^\text{scalar}$, and the channel-wise inner product between $h_\text{vec}^\text{node}$ and $g_\text{vec}^\text{node}$.
\end{enumerate}
In our experiments we used $64$ channels for all scalar and vectorial features.
The final readout is the sum of scalar node features over all nodes and channels. The cutoff radius when building the graph was chosen to be two times the $\sigma$ parameter of the LJ-potential.

\end{document}